# Phenomenological model of anomalous magnon softening and damping in half-metallic manganites.

A. Solontsov[1,2,*]

[1]*N.L. Dukhov Research Institute for Automatics, Suschevskaya str., 22, 127055 Moscow, Russia,*
[2]*State Center for Condensed Matter Physics, M. Zakharova str., 6-3, 115569 Moscow, Russia*

Dated: April 7, 2013

To describe anomalous zone-boundary softening and damping of magnons in manganites we present a phenomenological two-fluid model containing ferromagnetic Fermi-liquid and non-Fermi-liquid components. The Fermi-liquid component accounts for softening of zone-boundary magnons and for the Landau damping of magnons in the Stoner continuum arising at low frequencies due to zero-point effects. Coupling of the Fermi-liquid and non-Fermi-liquid fluids yields conventional long wavelength magnons damped due to their coupling with longitudinal spin fluctuations.

Metallic ferromagnetic manganites with colossal magnetoresistance besides numerous potential applications demonstrate fascinating physics including their anomalous magnetic dynamics below the Curie temperature $T_C$. In the high-$T_C$ manganites like $La_{0.7}Sr_{0.3}MnO_3$ ($T_C = 378$ K) and $La_{0.7}Pb_{0.3}MnO_3$ ($T_C = 355$ K) magnon spectrum can be well explained within the effective Heisenberg model[1], which for manganites follow from the canonical Vonsovsky and Zener description usually addressed as the double-exchange (DE) model[2].

In the low-$T_C$ manganites $Pr_{0.63}Sr_{0.37}MnO_3$ ($T_C = 301$ K), $La_{0.7}Ca_{0.3}MnO_3$ ($T_C = 238$ K), and $Nd_{0.7}Sr_{0.3}MnO_3$ ($T_C = 198$ K) magnons exhibit appreciate damping in the long wavelength limit and on approaching the Brillouin zone boundary they are softened and strongly damped[3], which cannot be attributed to the Heisenberg-type interactions. According to Ref. 3 at low temperatures magnon damping in these systems abruptly increases near the wavevector $k_c \sim 0.3$ (in the reciprocal lattice units, r.l.u.) where magnons merge with longitudinal optical phonons. However, no signs of magnon-phonon coupling could be seen in the magnon spectra including an energy gap between magnon and phonon frequencies. For the wavevectors $k_c > 0.3$ the magnon dispersion curves in the [001] and [110] directions follow almost flat optical phonon dispersions, and higher-frequency magnons are wiped out[1,3]. Probably, it would be more adequate to speak about locking of magnon modes on the optical phonon branches and about magneto-vibrational nature of the zone-boundary magnons[4].

Several mechanisms were proposed to account for the anomalies of zone-boundary magnon spectra in the low-$T_C$ manganites using the effective Heisenberg Hamiltonian or the DE approach[2] to describe i) four-magnon scattering[1], ii) magnon-phonon scattering with emission (absorption) of a phonon by magnons[5,6], iii) two-magnon scattering by itinerant electrons without spin-flip[7], iv) scattering of magnons by orbital fluctuations[8]. An additional mechanism due to v) scattering of magnons by longitudinal spin fluctuations (SF) with emission (absorption) of a SF by magnons first discussed in Ref. 9 could also describe magnon anomalies in manganites and will be discussed elsewhere. Here we only apply the latter mechanism to account for the long wavelength magnon damping.

First, we comment on the scattering mechanisms i)-iii) which result in magnon damping with the model-independent temperature and wavevector dependencies (model dependent parameters only affect the coupling matrix elements). Magnon damping due to four-magnon scattering processes i) were analyzed in the Heisenberg magnets[10,11] and ferromagnets with itinerant electrons[12] more than half a century ago and was shown to be negligibly small $\sim \mathbf{k}^2 T^2 \ln^2[k_B T / \hbar \omega_m(\mathbf{k})]_{T \to 0} \to 0$ at low temperatures ($k_B T \ll \hbar \omega_m(\mathbf{k})$), vanishing in the ground state ($T=0$), where $\mathbf{k}$ and $\omega_m(\mathbf{k})$ are the wavevector and magnon frequency, $T$ is the temperature. So, four-magnon scattering cannot contribute to the observed anomalous magnon damping at low temperatures in manganites.

Damping of magnons due to ii) magnon-phonon scattering processes was also discussed long ago for the Heisenberg magnets[13] and itinerant electron systems[14,15] and was shown to be exponentially small at low temperatures[14,15] $\sim \mathbf{k}^4 / T[\exp(k_B \vartheta_D^2 / \hbar \omega_m(\mathbf{k}) T)]_{T \to 0} \to 0$ (where $\vartheta_D$ is the Debye temperature), vanishing in the ground state. Three decades later this result was independently reexamined with respect to manganites using the DE[5] and Heisenberg[6] models. However, the authors[5,6] analyzing this mechanism arrive at absolutely different results, e.g., they find finite magnon damping in the ground state, which is in the strong disagreement with the previous findings[13-15]. Besides, it was reported finite magnon softening in the ground state[6]. These results are questionable and should be reexamined[16].

Two-magnon scattering by itinerant electrons without spin-flip iii) was first analyzed about five decades ago and shown to give finite magnon damping in the ground state of ferromagnets with itinerant electrons[17] $\sim \mathbf{k}^6$. Later this result was generalized to account for finite temperatures and complicated Fermi surfaces[18] and applied for the s-d-model for ferromagnets[19]. Independently similar results were obtained within the DE-model for manganites[7].

A joint feature of the approaches[7,8] describing the effects of scattering of magnons by electrons and orbital fluctuations is the continuous increase of magnon damping on approaching the Brillouin zone boundaries which disagrees with the observed abrupt rise in magnon damping near[1,3] $k \approx k_c \sim 0.3$ r.l.u. Finally, all mentioned mechanisms i)-iv) are based on the assumption of weak magnon damping although they are used to explain strong damping of magnons on approaching the Brillouin zone boundaries.

On the other hand, the obvious reason for the abrupt increase of magnon damping could be related to a possible intersection of the magnon dispersion curve with the continuum of Stoner excitations[20] leading to the strong Landau damping. However, Landau damping is ruled out if to use the canonical DE description of manganites or band structure calculations leading to their half-metallic character[21].

Here it is necessary to comment on the nature of half-metallic behavior of manganites. According to the mean field treatment of the one band DE exchange model itinerant $e_g$ electrons of Mn are 100% polarized and occupy only majority-spin subband due to strong Hund's-rule coupling. Within this treatment manganites have only one spin channel for the electron conductivity at the Fermi surface and should be considered as half-metallic. Indeed, spectroscopic measurements in $La_{0.7}Sr_{0.3}MnO_3$ ($T_C = 378$ K) using spin-polarized photoemission[22] at $T = 40$ K and inverse photoemission[23] at $T = 100$ K discovered 100% polarization which was questioned in Ref. 24. On the other hand, methods involving tunnel junctions give an essentially less polarization[25,26] $P \sim 80\%$.

Besides that, there are fundamental limitations for the half metallic behavior caused by mixing of electronic spins due to the effects of magnons and phonons coupled to the magnetic system and appearance of minority spin electrons at the Fermi surface at finite temperatures[24]. This conclusion was generalized by accounting for spin fluctuations in the classical approximation, which were shown to lead to a significant decrease of the polarization above $T > 0.4T_C$ and disappearance of the half-metallicity above[27] $T > 0.7T_C$. The main finding of Refs. 24 and 27 is that half-metallicity in manganites may be at the best expected at zero temperature. However, this ignores quantum effects of zero-point excitations which mix electronic spins and decrease spin polarization in the ground state. In manganites zero-point effects are mainly caused by longitudinal SF in the continuum of electron-hole pairs with the majority spin polarization, which yields the following rough estimation for the polarization[24,27] $P \sim M_S / M_0$, where[28,29] $M_S \approx [(4-x)^2 - 3m_{z.p.}^2]^{1/2} \mu_B$ and $M_0 \approx (4-x)\mu_B$ are the values of spontaneous magnetization per unit cell with and without account of SF, $m_{z.p.}^2$ is the squared amplitude of longitudinal zero-point SF,[28] and $x$ describes manganite doping. To estimate $m_{z.p.}^2$ we use the recently measured by inelastic neutron scattering squared amplitude of zero-point spin fluctuations in iron pnictides[30] $m_{z.p.}^2 \approx 3.2\mu_B^2$ (this value is surprisingly similar in most of the strongly correlated electron systems), which for half-metallic manganites gives $m_{z.p.}^2 \sim 0.53\mu_B^2$. This allows us to estimate the ground state polarization of manganites with $x = 0.3$ doping, $P \approx 0.94$. In real samples used in neutron scattering measurements the polarization may be less due to the effects of surfaces, interfaces, and structural inhomogeneties[24] which may lead to the observed value[25,26] ~80%. All this breaks down the concept of the half-metallic character of manganites even in the ground state and allow for the low-frequency Stoner excitations which may play an essential role in their magnetic dynamics and give rise to the strong Landau damping of magnons.

In the present Letter we argue that the observed effects of an abrupt increase of magnon damping at wavevectors $k_c > 0.3$ r.l.u. accompanied by their softening in manganites have a magnetic origin and can be well explained in the mean-field approximation by coupling of magnons with Stoner excitations arising due to the destruction of half-metallic ground state of manganites. On the other hand, the observed sudden increase of magnon damping in manganites may be viewed as a signature of Stoner excitations and non-half-metallicity of the ground state of manganites.

To analyze the transverse magnetic dynamics and anomalies of the magnon spectra of manganites we use a phenomenological approach based on the concept of the generalized magnetic susceptibility $\chi(\mathbf{k},\omega)$ which is associated with the transverse fluctuations of the magnetic order parameter with an amplitude $m(\mathbf{k},\omega)$, where $\omega$ is the frequency of SF. This approach is based on the assumption that all variables other than magnetic (e.g., individual Fermi excitations, charge and lattice fluctuations) are integrated out,[31,32] which means that all non-magnetic collective modes including lattice vibrations adiabatically follow magnetic (spin) fluctuations which can be viewed then as coupled spin-lattice fluctuations[31]. The approach is advantageous for manganites where spin-lattice coupling is believed to be strong[1].

To calculate the susceptibility $\chi(\mathbf{k},\omega)$ we use a phenomenological two-fluid model decoupling the amplitude $m(\mathbf{k},\omega) = m_1(\mathbf{k},\omega) + m_2(\mathbf{k},\omega)$ into two transverse components $m_1(\mathbf{k},\omega)$ and $m_2(\mathbf{k},\omega)$. The first one describes a ferromagnetic Fermi-liquid and is giving rise to "bare" magnons which are propagating in the long wavelength region ($k < k_c$) and strongly damped in the Stoner continuum ($k > k_c$). The second is related to a non-Fermi-liquid component of the magnetization. Both components are associated with the "partial" dynamical susceptibilities $\chi_{1,2}(\mathbf{k},\omega)$ of two fluids and describe their linear response $m_{1,2}(\mathbf{k},\omega) = \chi_{1,2}(\mathbf{k},\omega)B(\mathbf{k},\omega)$ to the transverse magnetic field $B(\mathbf{k},\omega)$.

Allowing for the coupling of two fluids one arrives at the following explicit form for the generalized dynamical susceptibility of the total system in the mean-field approximation (cf. Ref. 33)

$$\chi(\mathbf{k},\omega) = \frac{\chi_1(\mathbf{k},\omega) + \chi_2(\mathbf{k},\omega) + 2\lambda\chi_1(\mathbf{k},\omega)\chi_2(\mathbf{k},\omega)}{1 - \lambda^2 \chi_1(\mathbf{k},\omega)\chi_2(\mathbf{k},\omega)}. \quad (1)$$

Here $\lambda$ accounts for coupling of fluids which we assume to be ferromagnetic and neglect its spatial and time dispersions, $\lambda = const. > 0$. The susceptibility (1) founds the basis for the presented here phenomenological two-fluid model for transverse magnetic dynamics of manganites. Similar models were used for the description of the neutron scattering spectra in heavy fermion systems[33,34] and for the interpretation of the density functional calculations in iron pnictides[35].

It should be noted that Eq.(1) formally gives the dynamical susceptibility in the DE model if to consider $\chi_1(\mathbf{k},\omega)$ and $\chi_2(\mathbf{k},\omega)$ as "partial" susceptibilities of itinerant and localized moments subsystems and $\lambda$ as a coupling constant describing interaction between them[36]. Unlike the DE model the susceptibility (1) is not related to any microscopic Hamiltonian and founds the basis for a phenomenological approach being an alternative to microscopic descriptions, which is justified below by the analysis of the anomalies of the magnon spectra in manganites.

Here we use a minimal phenomenological model in order to explain anomalies of the magnon spectrum of manganites and take the partial Fermi-liquid susceptibility $\chi_1(\mathbf{k},\omega)$ in the following form,

$$\chi_1(\mathbf{k},\omega) = \chi_1(\mathbf{k}) \frac{\omega_0(\mathbf{k})}{\omega_0(\mathbf{k}) - \omega - i\Gamma(\mathbf{k},\omega)}, \quad (2)$$

where $\omega_0(\mathbf{k})$ is the frequency of "bare" optical magnons not vanishing in the long wavelength limit $(\omega_0(\mathbf{k}=0) = \omega_0 \neq 0)$. We emphasize that these "bare" magnons describing the poles of the "partial" magnetic susceptibility (2) are not the normal magnon modes of the total system and due to strong spin-lattice coupling in manganites they must be better viewed as magneto-vibrational modes.[14] We emphasize that these "bare" magnons are defined by the pole of the Here

$$\Gamma(\mathbf{k},\omega) = \theta[\omega - \omega_S(\mathbf{k})]\frac{\omega\omega_0(\mathbf{k})}{\omega_{fl}(\mathbf{k})} \quad (3)$$

is a conventional Fermi-liquid term accounting for the Landau damping in the Stoner continuum, $\omega_{fl}(\mathbf{k})$ is the characteristic frequency of transverse SF, the function $\theta[\omega - \omega_S(\mathbf{k})]$ is unity inside the Stoner continuum (when $\omega > \omega_S(\mathbf{k})$) and zero otherwise, and $\omega_S(\mathbf{k})$ is the lower boundary of the Stoner continuum. As we shall see later, $\omega_0(\mathbf{k})$ appears to be close to the frequencies of longitudinal optical phonons, confirming strong spin-lattice coupling in the low-$T_C$ manganites.

For the non-Fermi-liquid component of the magnetic susceptibility we use a static approximation, $\chi_2(\mathbf{k},\omega) \approx \chi_2(\mathbf{k})$. Here we do not consider the magnetic relaxation aside the Stoner continuum which we account for later.

The static partial susceptibilities $\chi_{1,2}(\mathbf{k})$ we present in the form

$$\chi_{1,2}^{-1}(\mathbf{k}) = \chi_{1,2}^{-1}[1 + a_{1,2}(\mathbf{k})], \qquad (4)$$

where the terms $a_{1,2}(\mathbf{k})$ account for their spatial dispersion which in the long wavelength limit we assume to be isotropic and quadratically dependent on the wavevector,

$$a_{1,2}(\mathbf{k}) = (\xi_{1,2}\mathbf{k})^2, \qquad (5)$$

where $\xi_{1,2}$ are correlation lengths.

Using Eq. (2) and assuming that $\omega << \omega_{fl}(\mathbf{k})$ we present the generalized dynamical susceptibility (1) near the magnon dispersion in the following explicit form

$$\chi(\mathbf{k},\omega) = \chi(\mathbf{k})z(\mathbf{k})\frac{\omega_m(\mathbf{k})}{\omega_m(\mathbf{k}) - \omega - i\tau^{-1}(\mathbf{k})}, \qquad (6)$$

which has a pole at the magnon frequency

$$\omega_m(\mathbf{k}) = \omega_0(\mathbf{k})[1 - \lambda^2\chi_1(\mathbf{k})\chi_2(\mathbf{k})], \qquad (7)$$

describing is a true normal mode of our two-fluid system. Here

$$\chi(\mathbf{k}) = \frac{\chi_1(\mathbf{k}) + \chi_2(\mathbf{k}) + 2\lambda\chi_1(\mathbf{k})\chi_2(\mathbf{k})}{1 - \lambda^2\chi_1(\mathbf{k})\chi_2(\mathbf{k})} \qquad (8)$$

is the static transverse susceptibility,

$$z(\mathbf{k}) = \chi_1(\mathbf{k})\frac{[1 + 2\lambda\chi_2(\mathbf{k})]^2}{\chi_1(\mathbf{k}) + \chi_2(\mathbf{k}) + 2\lambda\chi_1(\mathbf{k})\chi_2(\mathbf{k})}, \qquad (9)$$

is the relative weight of the magnon mode, and $\tau^{-1}(\mathbf{k}) = \Gamma(\mathbf{k},\omega_m(\mathbf{k}))$ describes damping of magnons in the Stoner continuum.

The factor $[1 - \lambda^2\chi_1(\mathbf{k})\chi_2(\mathbf{k})]$ plays the role of an enhancement factor for the susceptibility (8), and vanishes in the long wavelength limit,

$$1 - \lambda^2 \chi_1 \chi_2 = 0. \tag{10}$$

This equality could plays a role of the magnetic equation of state if one would know the dependencies of $\chi_{1,2}$ on the magnetic order parameter.

The important consequence of Eq.(10) is the gapless character of the magnon spectrum (7) which with account of Eq.(4) takes the form

$$\omega_m(\mathbf{k}) = \omega_0(\mathbf{k}) \frac{a_1(\mathbf{k}) + a_2(\mathbf{k}) + a_1(\mathbf{k})a_2(\mathbf{k})}{[1+a_1(\mathbf{k})][1+a_2(\mathbf{k})]}, \tag{11}$$

and vanishes in the long wavelength limit

$$a_{1,2}(\mathbf{k}) \approx (\xi_{1,2}\mathbf{k})^2 \ll 1, \tag{12}$$

where the magnon dispersion is isotropic and quadratic

$$\omega_m(\mathbf{k}) \approx \omega_0(\mathbf{k})[(\xi_1\mathbf{k})^2 + (\xi_2\mathbf{k})^2] \approx D\mathbf{k}^2 \tag{13}$$

and $D = \omega_0(\mathbf{k}=0)(\xi_1^2 + \xi_2^2)$ is the magnon stiffness. Eq.(13) agrees with the measured isotropic long wave-vector spectrum of manganites[1]. As it follows from Eq. (9) the weight factor $z(\mathbf{k})$ in this limit equals unity ($z(\mathbf{k} \to 0) \approx 1$).

It should be noted that within our two-fluid model the high-frequency mode $\omega = \omega_0(\mathbf{k})$ of the Fermi-liquid fluid is transferred into the acoustical magnon (11) due the exchange coupling of the fluids. In a sense, this is similar to the transfer of the plasma ion mode in metals into the acoustical longitudinal phonon due to the electron Coulomb screening of ions.

The magnon spectrum and damping given by Eqs. (6) and (11) may account for anomalous magnon softening and damping on approaching the Brillouin zone boundaries observed in manganites[1,3]. To explain the anomalies of the magnon spect**r**um in manganites within our phenomenological model one should take into account anisotropic dependencies of $\omega_0(\mathbf{k})$, $\omega_{fl}(\mathbf{k})$, and $a_{1,2}(\mathbf{k})$ on the wavevector, which must be inferred from the measured data.

Due to the lack of the experimental data for the magnon spectra in the whole Brillouin zone we shall limit ourselves to the qualitative analysis of magnon anomalies near the Brillouin zone boundaries along several directions. Namely, we analyze magnon spectrum along the [001] and [110] directions in the low-$T_C$ manganites $La_{0.7}Ca_{0.3}MnO_3$, $Pr_{0.63}Sr_{0.37}MnO_3$, and $Nd_{0.7}Sr_{0.3}MnO_3$.

This allows us to make further approximations minimizing the number of parameters that should be taken from experiments. Namely, we neglect the spatial dispersion of the frequency $\omega_{fl}(\mathbf{k}) \approx \omega_{fl}(0) = \omega_{fl}$ and the partial susceptibility $\chi_1(\mathbf{k}) \approx \chi_1$ (or set $a_1(\mathbf{k}) = 0$ and $\xi_1 = 0$), and use an approximate equality $a_2(\mathbf{k}) \approx (\xi_2 \mathbf{k})^2$ throughout the Brillouin zone.

Then the magnon frequency (11) is given by the following explicit formula

$$\omega_m(\mathbf{k}) = \omega_0(\mathbf{k}) \frac{(\xi_2 \mathbf{k})^2}{1 + (\xi_2 \mathbf{k})^2}. \tag{14}$$

It has a quadratic dispersion in the long wavelength limit $(\xi_2 \mathbf{k})^2 \ll 1$ with the magnon stiffness $D = \omega_0 \xi_2^2$, softens at $|\mathbf{k}| \sim \xi_2^{-1} = k_s$ and saturates $\omega_m(\mathbf{k}) \approx \omega_0(\mathbf{k})$ at short wavelengths $(\xi_2 \mathbf{k})^2 \gg 1$.

Magnon damping in Eq.(6)

$$\tau^{-1}(\mathbf{k}) = \frac{\omega_0^2(\mathbf{k})}{\omega_{fl}} \theta[\omega_m(\mathbf{k}) - \omega_S(\mathbf{k})] \tag{15}$$

exhibits a discontinuous jump when the magnon dispersion curve crosses the Stoner continuum boundary at the wavevector $k = k_c$ defined by

$$\omega_m(\mathbf{k}) = \omega_S(\mathbf{k}), \tag{16}$$

In microscopic descriptions magnon dispersion possesses logarithmic softening when crosses the Stoner continuum boundary[36], so, one should expect that the wavevector $k_c$ is close to the value $k_s$ at which the magnon dispersion demonstrate softening.

In the low-$T_C$ manganites $Pr_{0.63}Sr_{0.37}MnO_3$, $La_{0.7}Ca_{0.3}MnO_3$, and $Nd_{0.7}Sr_{0.3}MnO_3$ the magnon dispersions measured by inelastic neutron scattering[1] are quadratic up to the wavevector $k_s \sim 0.2$ r.l.u. At higher wavevectors magnon energies exhibit softening and saturation, which is strongly anisotropic being about 20 and 40 meV along [001] and [110], and are close to the energies of longitudinal optical phonons in these directions. It should be noted that in $Nd_{0.7}Sr_{0.3}MnO_3$ zone boundary magnons in the [110] direction are overdamped to be observed experimentally[3]. To account for the anisotropy of magnon softening and saturation in the low-$T_C$ manganites we are to allow for the wavevector dependence of the "bare" magnon frequency of the ferromagnetic fluid

$$\omega_0(\mathbf{k}) \approx \begin{cases} \omega_0, & (\xi_2 \mathbf{k})^2 \ll 1, \\ \omega_1, & (\xi_2 \mathbf{k})^2 \gg 1, [001] \\ \omega_2, & (\xi_2 \mathbf{k})^2 \gg 1, [110] \end{cases}, \tag{17}$$

where we take the saturation energies in the [001] and [110] directions $\hbar\omega_1 \approx 22$ and $\hbar\omega_2 \approx 40$ meV from Refs. 1 and 3, which appear to be close to the energies $\hbar\Omega_1 \approx 27$ and $\hbar\Omega_2 \approx 49$ meV of the optical longitudinal phonons[3].

It should be emphasized that the frequencies $\omega_{1,2}$ are the magnon (or magneto-vibrational) frequencies defined by the pole of the transverse magnetic susceptibility (6). Their proximity to the optical phonons frequencies $\Omega_{1,2}$ probably results from the spin-lattice couplings incorporated into magnetic dynamics of our model. This proximity in our phenomenological approach is not necessary for the description of magnon softening and broadening unlike the microscopic approach[5,6] assuming an intersection of magnon and phonon dispersions.

First, we analyze the long wavelength magnon spectrum in the low-$T_C$ manganites $Pr_{0.63}Sr_{0.37}MnO_3$, $La_{0.7}Ca_{0.3}MnO_3$, and $Nd_{0.7}Sr_{0.3}MnO_3$ basing on Eq.(11). To minimize the number of the parameters of our model we assume the energy $\hbar\omega_0$ in Eq.(17) to be equal to its minimal zone-boundary value $\hbar\omega_0 = \hbar\omega_1 \approx 22$ meV. Using the measured magnon stiffness $\hbar D = \hbar\omega_0 \xi_2^2 \approx 165$ meVA$^2$ one estimates the correlation length $\xi_2 \approx 2.74$ A and the wavevector $k_s = \xi_2^{-1} \approx 0.2$ r.l.u. which agrees with the measured vector where starts softening of the magnon spectra[1,3] and is close to the value $k_c \approx 0.3$ marking the abrupt increase of magnon damping in [001] and [110] directions. From the analysis of magnon damping in the [001] direction we can estimate the SF frequency $\omega_{fl}$ in the Stoner continuum. Using the equality $\tau^{-1} = \omega_1^2 / \omega_{fl}$ following from Eq.(15) and the measured abrupt increase of magnon damping from ~4 to 12 meV at $k_c \approx 0.3$ we get the estimate $\hbar\omega_{fl} \approx 60$ meV which looks reasonable for SF in itinerant electron magnets[36] and satisfies the assumption $\omega_{fl} >> \omega_m(\mathbf{k})$ we used in Eq.(6).

Up to now we analyzed magnon damping and softening on the basis of the mean-field magnetic susceptibility (6) which is the linear approximation to non-linear magnetic dynamics[32] accounting for mode-mode couplings and give rise to the mentioned above i)-v) scattering mechanisms.

Finally, we comment on the long wavelength magnon damping in the low-$T_C$ manganites which is also anomalously high and at the wavevector $k_s \approx 0.2$ r.l.u. is about 4 meV. The most effective mechanism of magnons in magnets with itinerant electrons (besides damping in the Stoner continuum) is caused by the scattering processes with emission (absorption) of longitudinal SF by magnons which was first proposed in Ref. 37. In the Born approximation it gives the following explicit expression for damping of low-temperature ($k_B T \ll \omega_m(\mathbf{k})$) long wavelength magnons[9]

$$\tau^{-1}(\mathbf{k},T) = \frac{4}{5\mu} \frac{\omega_m^2(\mathbf{k})}{\omega_{SF}} \sim \mathbf{k}^4 \qquad (18)$$

where $\mu$ is the magnetic moment of the unit cell in the units of the Bohr magneton and $\omega_{SF}$ is the characteristic frequency of longitudinal SF (different from the analogous frequency $\omega_{fl}$ of transverse SF in the Stoner continuum). Comparing magnon damping (18) with the inverse lifetime due to, e.g., magnon-electron scattering[18] one finds that the latter contains an additional small factor $\sim \mathbf{k}^2$ making magnon-electron scattering mechanism less effective at long wavelengths.

Eq.(18) is in reasonable agreement with the wavevector dependence of long wavelength damping in the low-$T_C$ manganites[1,3]. Using the values[3] $\hbar\tau^{-1} \approx 4$ meV and $\hbar\omega_m \approx 14$ meV for magnons at the wavevector $k_c \approx 0.3$ r.l.u. in the [001] direction and the magnetic moment $\mu \approx 3.7$, we find the energy of longitudinal SF $\hbar\omega_{SF} \approx 26$ meV. The difference between energies $\hbar\omega_{fl} \sim \chi_1^{-1}$ and $\hbar\omega_{SF} \sim \chi_l^{-1}$ of the transverse and longitudinal SF proportional to the appropriate inverse magnetic susceptibilities[37] should be due to the exchange enhancement of the longitudinal susceptibility $\chi_l$ compared to the "partial" transverse one $\chi_1$, $\omega_{fl}/\omega_{SF} \sim \chi_l/\chi_1 \approx 2.3$, which looks reasonable.

To conclude, our main finding is that zone-boundary magnon anomalies in the low-$T_C$ manganites cannot be understood without using a concept of the Stoner excitations which should arise even in the ground state of half-metallic ferromagnets due to mixing of electronic spins by zero-point longitudinal spin fluctuations. We also argue that softening of magnons near the Brillouin zone boundaries is closely related to their magneto-vibrational character and in our minimal description can be well described within a two-fluid model containing a ferromagnetic Fermi liquid and non-Fermi-liquid components. Aside the Stoner continuum magnon damping can be described by various mode-mode scattering processes among which processes with emission (absorption) of longitudinal SF by magnons are the most important.

The work was supported by the State Atomic Energy Corporation of Russia "ROSATOM".

* *E-mail address*: asolontsov@mail.ru.